\documentclass[twocolumn,showpacs]{revtex4}

\usepackage{graphicx}
\usepackage{dcolumn}
\usepackage{bm}
\newcommand{\beq}{\begin{equation}}
\newcommand{\eeq}{\end{equation}}
\newcommand{\bea}{\begin{eqnarray}}
\newcommand{\eea}{\end{eqnarray}}
\newcommand{\bec}{\begin{center}}
\newcommand{\enc}{\end{center}}

\newcommand{\om}{\omega}
\newcommand{\Om}{\Omega}
\newcommand{\eps}{\epsilon}
\newcommand{\rmd}{{\rm d}}
\newcommand{\rmg}{{\rm g}}
\newcommand{\rmi}{{\rm i}}
\newcommand{\rmx}{{\rm x}}

\newcommand{\veck}{\bm{k}}
\newcommand{\vecq}{\bm{q}}

\begin{document}
\title{Quantum Zeno Effect for Exponentially Decaying Systems}
\author{Kazuki Koshino}
\email{ikuzak@aria.mp.es.osaka-u.ac.jp}
\affiliation{
Core Research for Evolutional Science and Technology (CREST),
Japan Science and Technology Corporation (JST),
c/o Department of Physical Science, Graduate School of Engineering Science,
Osaka University, Toyonaka, Osaka 560-8531, Japan
}
\author{Akira Shimizu}
\email{shmz@ASone.c.u-tokyo.ac.jp}
\affiliation{
Department of Basic Science, University of Tokyo, 
3-8-1 Komaba, Tokyo 153-8902, Japan
}
\date{\today}
\begin{abstract}

The quantum Zeno effect -- suppression of decay by frequent measurements 
-- was believed to occur only when the response of the detector is so quick 
that the initial tiny deviation from the exponential decay law is detectable. 
However, we show that it can occur even for exactly exponentially 
decaying systems, for which this condition is never satisfied, 
by considering a realistic case 
where the detector has a finite energy band of detection. 
The conventional theories correspond to the limit of an infinite bandwidth. 
This implies that the Zeno effect occurs more widely than expected so far.

\end{abstract}
\pacs{03.65.Xp,06.20.Dk,03.65.Yz}
\maketitle

The survival probability $s(t)$ of an unstable quantum system
generally decreases quadratically with time $t$
at the beginning of the decay,
and later follows the well-known exponentially decay law.
The crossover takes place at $t\sim\tau_j$,
which is called the jump time.
Noticing this fact and simply applying the 
projection postulate on measurements, 
it was predicted that frequent measurements with time intervals $\tau$ 
result in the suppression of the decay
if $\tau$ is small enough to satisfy $\tau\lesssim\tau_j$ 
\cite{MS}.
Similarly, the suppression was supposed to occur 
by continuous measurement using an apparatus with short response time $\tau$
satisfying $\tau\lesssim\tau_j$ \cite{Sch}.
This prediction, called the quantum Zeno effect (QZE),
is one of the most curious results of the quantum measurement,
and has been attracting much attention 
both theoretically
\cite{MS,Peres,Panov,Sch,nature,FNP,KS} 
and experimentally
\cite{Itano,Knight,QZE_coll,laser_exp2,FGR}.
However, in most unstable systems,
$\tau_j$ is extremely small so that  
the condition $\tau\lesssim\tau_j$ cannot be satisfied 
by existing detectors.
It has been therefore believed that the QZE
would not occur in most unstable systems.

However, these conventional arguments on the QZE need to be reconsidered,
because, in general, real measurement processes 
are not `projective measurements' that are described by the 
projection postulate \cite{NC,SF,QNoise}. 
A more general and accurate way of analyzing measurement processes is 
to apply the laws of quantum theory to a larger system
which include {\em both} the original quantum system to be 
measured {\em and} a part of measuring apparatus 
\cite{SF,QNoise}.
Such an approach has been successfully applied, e.g., to
quantum optics 
\cite{QNoise,Mandel,Gla,QND}
and the QZE \cite{Sch,KS}.
This enables one to study 
not only the temporal evolution of the unstable system
but also the response of the measuring apparatus.
Moreover, one can discuss general measurement processes
such as measurement with a finite probability of error 
\cite{KS,QND}.

In this paper, we apply this modern approach to analyze the QZE 
in a realistic situation 
where the unstable system is monitored continuously using
a measuring apparatus (detector) with a 
finite energy band of detection.
Furthermore, we consider the case of `indirect measurement,'
in which the detector acts only on the decay products, 
because some of the previous works assuming direct interactions 
have often been criticized as not being a genuine QZE.
We show that the QZE can take place 
even in systems that exactly follow the exponential decay law,
for which $\tau_j$ is infinitesimal hence the conventional condition 
for the QZE, $\tau\lesssim\tau_j$, is never satisfied.
We clarify its physical origin, derive
the conditions for inducing the QZE in such systems, 
and show that the conventional projection-based theory can be reproduced 
in the ideal, but unrealistic limit of an infinite detection bandwidth.
Our results imply that 
the response time required for the QZE is not so short, 
the required jump time is not so long, 
and hence the QZE may take place much more widely
than expected so far.

As the unstable quantum system
we consider the excited state of a two-level atom,
which decays to the ground state with a finite lifetime
accompanying emission of a photon.
By detecting the emitted photon, 
the observer can know the decay of the atom.
As the photon detection process,
we consider a standard one \cite{QNoise,Mandel,Gla}:
In the photodetector, the emitted photon is absorbed by 
a semiconductor and an electron-hole pair is generated.
It finally yields a macroscopic signal after magnification processes,
by which the observer knows the decay of the excited atom.
Following the quantum theory of photon counting
\cite{QNoise,Gla,Mandel},
we treat the relevant part of the detector, i.e., 
electron-hole pairs, 
as a part of the total quantum system.
The electron-hole pairs can be described as
bosonic elementary excitations,
because their density is low in the detection process.
The total quantum system is therefore composed of three parts:
(i) A two-level atom with the ground state $|g\rangle$
and the excited state $|x\rangle$ with the transition energy $\Om$
(we hereafter take $\hbar =c=1$),
(ii) the photon field,
whose eigenmodes are labeled by the wavevector $\veck$,
and (iii) the field of the electron-hole bosonic excitation in the detector.
By taking the energy of $|g\rangle$ as the origin of the energy,
the Hamiltonian of the total system is given as follows:
\bea
{\cal H}&=&{\cal H}_0+{\cal H}_1+{\cal H}_2,
\label{eq:H}
\\
{\cal H}_0 \! &=& \! \Om |\rmx\rangle\langle\rmx|,
\label{eq:H0}
\\
{\cal H}_1 \! &=& \! 
\int\rmd \veck
\left[
\left(\xi_{\veck}|x\rangle\langle g|b_{\veck}+{\rm H.c.}\right)
+k\ b_{\veck}^{\dagger}b_{\veck}
\right],
\label{eq:H1}
\\
{\cal H}_2 \! &=& \! 
\int \! \! \int \! \rmd\veck \, \rmd\om 
\left[
\left(\sqrt{\eta_{\veck}}b_{\veck}^{\dagger}c_{\veck\om}+{\rm H.c.}\right)
+\om c_{\veck\om}^{\dagger}c_{\veck\om}
\right].
\label{eq:H2}
\eea
Here, ${\cal H}_1$ represents the atom-photon interaction,
where $b_{\veck}$ is the annihilation operator 
for the photon with wavevector $\veck$,
whose energy is $k=|\veck|$.
The atom-photon part ${\cal H}_0+{\cal H}_1$
constitutes the original unstable quantum system to be measured.
${\cal H}_2$ represents the interaction between the emitted photon and the detector.
Every photon mode is coupled to a continuum of 
the bosonic elementary excitations in the detector,
whose annihilation operator is denoted by $c_{\veck\om}$.
The commutation relations are orthonormalized as 
$[b_{\veck},b^{\dagger}_{\veck'}]=\delta(\veck-\veck')$ and 
$[c_{\veck\om},c^{\dagger}_{\veck'\om'}]=\delta(\veck-\veck')\delta(\om-\om')$.

In order to demonstrate that 
the QZE does occur even when $\tau\gtrsim\tau_j$, 
for which the QZE cannot be expected according to the conventional theories,
we study the limiting case of infinitesimal $\tau_j$ 
\cite{finite-tauj}.
Namely, we investigate the case where 
the survival probability $s(t)$ of the initial excited state 
exactly follows the exponential decay law, $s(t)=\exp(-\gamma t)$,
when the original system is not measured,
i.e., when ${\cal H}_2=0$.
In the present model, we obtain such an exact exponential decay 
for the original quantum system (described by ${\cal H}_0+{\cal H}_1$)
by putting the atom-photon coupling 
after angular integration
to be independent of the photon energy $k$;
\beq
\int\rmd\vecq\ |\xi_{\vecq}|^2\delta(|\vecq|-k) \equiv 
\frac{\gamma}{2\pi},
\eeq
and by extending the lower limit of 
the photon energy to $-\infty$. 
In this case, $\tau_j$ becomes infinitesimal, and
$\tau > \tau_j$ for any finite $\tau$.
We show in the following that the QZE can occur 
even in this case \cite{finite-tauj}.
The point is that any real detector has a finite bandwidth of detection.
In contrast, the use of the projection hypothesis corresponds to 
the use of an ideal but unrealistic detector that
has an infinite bandwidth.

To demonstrate this, we here take 
the following simple form for $\eta_{\veck}$;
\beq
\eta_{\veck}=\eta_k\equiv
\frac{\eta/2\pi}{1+[(k-\Omega)/\Delta]^n},
\eeq
where $n$ is a large even integer
(we take $n=6$ in the numerical examples).
In this case, 
photons within the energy range $|k-\Omega| \lesssim \Delta$ 
are counted 
by the detector with a timescale of $\eta^{-1}\equiv\tau$,
while photons outside of this energy range
are not counted.
Thus, the detector has a detection band
centered around the atomic transition energy $\Om$
with the bandwidth $\simeq 2\Delta$.
The atomic decay is continuously monitored with this detector.

\begin{figure}
\includegraphics{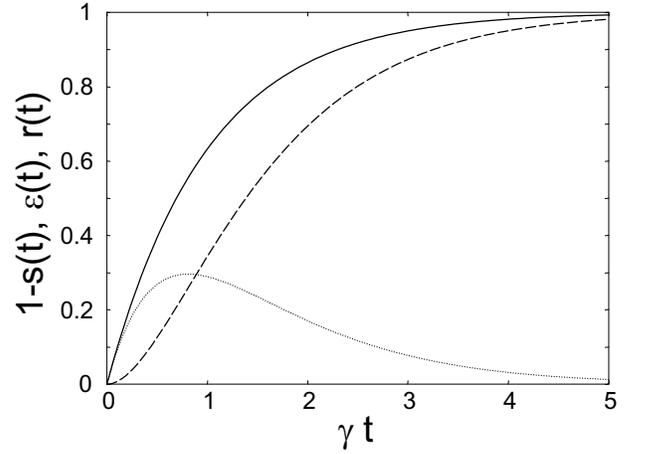}
\caption{\label{fig:1}
Temporal evolution of $1-s(t)$ (solid curve),
$\eps(t)$ (dotted), and $r(t)$ (broken).
The parameters are chosen as $2\pi\Delta/\gamma=100$ and $\eta/\gamma=1.5$.
It is seen that $r(t)$ follows $1-s(t)$ with a delay time
about $\tau \equiv \eta^{-1}$. 
}\end{figure}

To see the temporal evolution of the whole system,
we put $|\psi(t)\rangle=e^{-\rmi{\cal H}t}|\rmx,0,0\rangle
=f(t)|\rmx,0,0\rangle+\int\rmd \veck\ f_{\veck}(t)|\rmg,\veck,0\rangle
+\int\int\rmd \veck\rmd\om\ f_{\veck\om}(t)|\rmg,0,\veck\om\rangle$,
and define three probabilities of physical interest;
$s(t)=|f(t)|^2$ (survival probability of the atom),
$\varepsilon(t)=\int\rmd \veck|f_{\veck}(t)|^2$ 
(probability that the atom has decayed but the emitted photon 
is not absorbed by the detector), and
$r(t)=\int\int\rmd \veck\rmd\om|f_{\veck\om}(t)|^2$ 
(probability that the emitted photon is absorbed).
$r(t)$ can be interpreted as 
the probability of getting a detector response,
whereas $\varepsilon(t)$ is the probability 
that the detector reports an erroneous result.
One of the advantages of the present theory is that all of 
these interesting quantities can be calculated.
A numerical example is shown in Fig.~\ref{fig:1}.
Since $\Delta \gg \gamma$ in this example, 
the atomic natural linewidth ($\sim\gamma$) 
is completely covered by the detection band ($\sim\Delta$),
and the emitted photon is counted almost perfectly
($r(t)\simeq 1$ for $t \to \infty$).
It is observed that $r(t)$ follows the decay probability $1-s(t)$ 
with a delay time $\simeq \tau$ ($= \eta^{-1}$).
Hence, $\tau$ is the response time of the detector.
To be more precise, $\tau$ is the {\em lower limit} of
the response time because
additional delays in the response, 
such as delays in signal magnification processes, 
may occur in practical experiments.
In discussing fundamental physics, 
the limiting value is more significant     
than practical values, which depend strongly  
on detailed experimental conditions.
A typical value of $\tau$ for GaAs is $10^{-15}$s, which is much shorter than
the practical response times of commercial photodetectors, which 
range from $10^{-6}$ to $10^{-13}$s.

\begin{figure}
\includegraphics{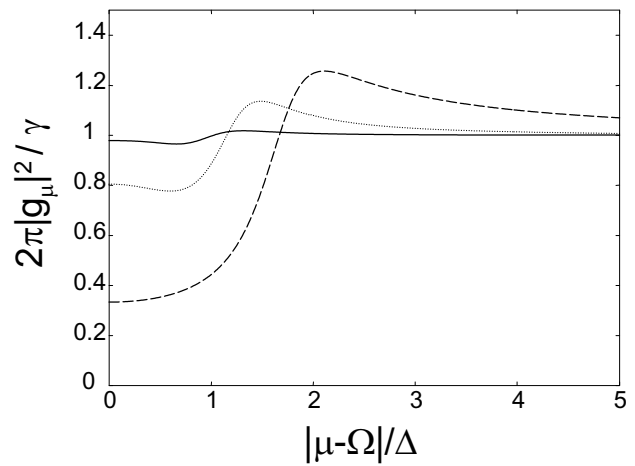}
\caption{\label{fig:2}
Plot of $|g_{\mu}|^2$.
$\eta/2\pi\Delta$ is chosen at 0.01 (solid curve), 0.1 (dotted), and 1 (broken).
}\end{figure}

In order to see the QZE, 
we investigate how the survival probability $s(t)$
is affected by the detector parameters such as 
the response time $\tau$ and 
the detection bandwidth $\Delta$ of the detector.
For this purpose, we transform ${\cal H}$ 
into the following renormalized form 
\cite{KS},
\bea
{\cal H}&=&{\cal H}_0+\bar{\cal H}_1+\bar{\cal H}_2,
\label{eq:Hnew}
\\
\bar{\cal H}_1 \! &=& \! 
\int\rmd\mu
\left[
\left(g_{\mu}|x\rangle\langle g|B_k+{\rm H.c.}\right)
+\mu\ B_{\mu}^{\dagger}B_{\mu}
\right],
\label{eq:H1new}
\eea
where $\bar{\cal H}_2$ is composed of terms which are 
decoupled from the atom.
Here, $B_{\mu}$ is a coupled-mode operator 
that is a linear combination 
of $b_{\veck}$'s and $c_{\veck\om}$'s 
\cite{KS},
which is orthonormalized in terms of a one-dimensional label $\mu$ as 
$[B_{\mu},B_{\mu'}^{\dagger}]=\delta(\mu-\mu')$, and 
$|g_{\mu}|^2$ is given by
\beq
|g_{\mu}|^2=\frac{\gamma}{2\pi}\int\rmd k
\frac{\eta_k}{|\mu-k-\rmi\pi\eta_k|^2},
\label{eq:g2}
\eeq
which we call the renormalized form factor.
Thus, the atom is coupled to a single continuum of $B_{\mu}$, 
with the renormalized form factor.
When the atom-photon system 
is free from the measurement, i.e., when $\eta=0$,
the form factor reduces to the `free' value, 
$|g_{\mu}|^2=\gamma/2\pi$,
and the atomic decay exactly follows the exponential law,  
$s(t)=\exp(-\gamma t)$.
When the detector is present,  
on the other hand, 
$|g_{\mu}|^2$ is modified (renormalized) as shown in 
Fig.~\ref{fig:2}, where $|g_{\mu}|^2$ is plotted 
for several values of $\eta/\Delta$.
It is seen that 
$|g_{\mu}|^2$ is decreased for $|\mu-\Om|\lesssim\Delta$
(inside the detection band)
and is increased for $|\mu-\Om|\gtrsim\Delta$ (outside).
For larger $|\mu-\Om|$,
modification of $|g_{\mu}|^2$ becomes small, 
approaching zero (i.e., $|g_{\mu}|^2 \to \gamma/2\pi$) 
for $|\mu-\Om|\gg\Delta$. These modifications take place 
satisfying a sum rule, $\int\rmd\mu|g_{\mu}|^2=$ constant,
which is derived from Eq.~(\ref{eq:g2}).
The figure also shows that the modification of $|g_{\mu}|^2$ 
occurs more significantly for larger $\eta/\Delta$.
This can be seen clearly in the limit of $n\rightarrow\infty$, 
where $|g_{\mu}|^2$ can be analytically evaluated 
and its value at the atomic transition energy is given by
$|g_{\Om}|^2=(\gamma/\pi^2)\arctan(2\Delta/\eta)$.
Thus, $|g_{\Om}|^2$ becomes significantly 
smaller than the free value $\gamma/2\pi$
when $\eta/\Delta\gtrsim 1$, i.e., when
\beq
\tau\cdot\Delta\lesssim 1.
\label{eq:cond2}
\eeq

\begin{figure}
\includegraphics{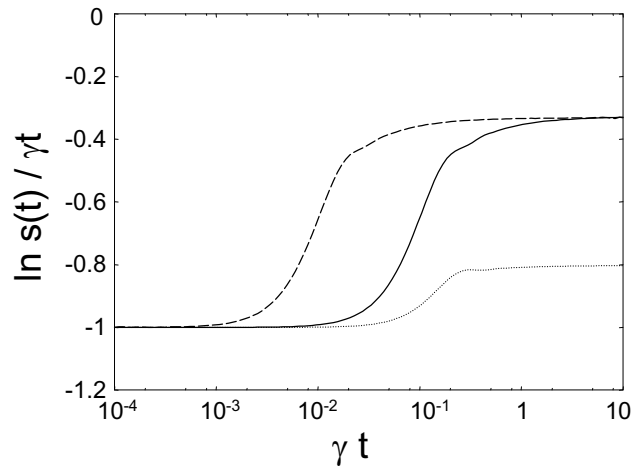}
\caption{\label{fig:3}
Plot of $\ln s(t)/(\gamma t)$, i.e.,
the time dependence of the decay rate.
$\{2\pi\Delta/\gamma,\eta/\gamma\}$ are chosen at
$\{100,100\}$ (solid curve),
$\{100,10\}$ (dotted), and
$\{1000,1000\}$ (broken)
The decay rate changes from the free rate $\gamma$
to the suppressed rate $2\pi|g_{\Om}|^2$ at $t\sim\Delta^{-1}$.
}\end{figure}

The survival probability $s(t)$ is governed
by the functional form of $|g_{\mu}|^2$.
In order to visualize the decay rate,
$\ln s(t)/(\gamma t)$ is plotted in Fig.~\ref{fig:3}.
The figure clarifies the following 
two-stage behavior of the atomic decay:
In the first stage ($t\lesssim\Delta^{-1}$),
the atom decays with the free decay rate $\gamma$,
while, in the second stage ($t \gtrsim \Delta^{-1}$),
the atom decays with a suppressed decay rate $2\pi|g_{\Om}|^2$.
For the values of parameters chosen in Fig.~\ref{fig:3}, 
the atom is kept almost undecayed ($s(t) \simeq 1$) in the first stage, 
and significant decay occurs in the second stage
with the suppressed decay rate.
Namely, the QZE surely takes place, 
although the system exhibits an exact
exponential decay in the absence of measurement.

The two-stage behavior can be understood 
with a help of the perturbation theory.
Applying the lowest-order perturbation 
to the renormalized form Eq.~(\ref{eq:Hnew}), 
we obtain the decay probability as
\beq
1-s(t)=\int\rmd\mu\ |g_{\mu}|^2
\frac{\sin^2[(\mu-\Om)t/2]}{[(\mu-\Om)/2]^2}.
\eeq
Taking into account that the main contribution in the integral 
comes from the region of $\mu$ satisfying 
$|\mu-\Om| \lesssim 2\pi t^{-1}$,
we evaluate the right-hand side in two cases:
In the case of $t \ll \Delta^{-1}$,
$|g_{\mu}|^2$ can be approximated by $|g_{\infty}|^2=\gamma/2\pi$,
which gives the free decay rate $1-s(t)=\gamma t$;
whereas in the opposite case of $t \gg \Delta^{-1}$,
$|g_{\mu}|^2$ can be approximated by $|g_{\Om}|^2$,
which gives the suppressed decay rate $1-s(t)=2\pi|g_{\Om}|^2t$.
Thus, the decay rate changes from the free rate  
to the suppressed rate at $t\sim\Delta^{-1}$.

The QZE occurs for exponentially decaying systems
when the following two conditions are satisfied:
(i) The transition from the first to the second stage 
should occur before the atom decays.
Since the survival probability at $t\sim\Delta^{-1}$
is given by $s(\Delta^{-1})\simeq\exp(-\gamma/\Delta)$,
this condition is expressed as
\beq
\gamma/\Delta\ll 1,\label{eq:cond1}
\eeq
which means that the detection band should completely cover 
the natural linewidth of the atom.
(ii) The decay rate in the second stage ($t\gtrsim\Delta^{-1}$)
should be significantly suppressed from the free decay rate.
This condition is expressed by inequality (\ref{eq:cond2}).

The latter condition explains why the QZE was not obtained 
for exponentially decaying systems
by theories based on the projection hypothesis.
By applying the projection operator, 
the quantum coherences between 
$|x,0\rangle$ and $|g,\veck\rangle$'s are destroyed
regardlessly of the energy of the emitted photon.
Therefore, the projection-based theory 
corresponds to  
$\Delta\rightarrow\infty$ \cite{cm:finite-tau}.
In such a limit, however, 
inequality (\ref{eq:cond2}) 
cannot be satisfied and the QZE never occurs.
Since $\Delta$ of any real detector is finite, 
such a limit is rather unphysical.

To conclude this paper, we make three remarks.
We have presented results for the case where
the transition energy $\Om$ is inside the detection band.
It is worth mentioning that
if the transition energy $\Om$ lies outside the detection band, 
the `anti QZE' is possible, which is the enhancement of the decay 
by the measurement
\cite{nature,FNP,KS}.
This may be understood from Fig.~\ref{fig:2}, which shows that 
$|g_\mu|^2 > \gamma/2\pi$ for $|\mu - \Omega| \gtrsim \Delta$.

We have assumed continuous measurement.
Regarding frequent discrete measurements, we note that 
real detectors have finite response times $\tau$, although 
the conventional theories on the QZE often assumed that 
each measurement was an instantaneous projective measurement.
Since each measurement takes a detector a time $\tau$,
the intervals of the measurements should be longer than $\tau$.
Therefore, our case of continuous measurement corresponds to 
the most frequent discrete measurements using the detector 
that is described by Eq.\ (\ref{eq:H2}).
In some cases, continuous measurement with the response time $\tau$
corresponds to frequent instantaneous measurements 
with the intervals of order $\tau$ \cite{Sch}.
However, it should be stressed that this does {\em not} mean that 
continuous measurement {\em always} corresponds to some 
frequent instantaneous projective measurements.
In fact, the result of this work is a counterexample.

Finally, we remark that the QZE is closely related 
to the cavity quantum electrodynamics (QED) \cite{CQED}.
In fact, Eq.\ (\ref{eq:H2}) is similar to
the equation describing the cavity QED
where $b_{\veck}$'s correspond to discretized cavity modes
while $c_{\veck\om}$'s correspond to leaky modes, 
or, when the cavity is composed of photoabsorptive materials,
to excitations in the material.
In principle, these quanta can be utilized for detection of the decay, 
if appropriate magnification (and, if necessary, detection) 
processes are followed. In such cases, 
the lower limits of the response time $\tau$ would be determined 
in terms of the material parameters such as $\eta_k$.
The present theory not only demonstrates the QZE for the 
exponentially decaying systems, but also
reveals that the cavity QED is related to the QZE 
through such implicit relations \cite{sensitive}.



\begin{thebibliography}{99}

\bibitem{MS}
B. Misra and E. C. G. Sudarshan,
J. Math. Phys. {\bf 18}, 756 (1977).

\bibitem{Sch}
L. S. Schulman,
Phys. Rev. A {\bf 57}, 1509 (1998).

\bibitem{Peres}
A. Peres and A. Ron, 
Phys. Rev. A {\bf 42}, 5720 (1990). 

\bibitem{Panov}
A. D. Panov,
Ann. Phys. {\bf 249}, 1 (1996).

\bibitem{nature}
A. G. Kofman and G. Kurizki,
Nature {\bf 405}, 546 (2000).

\bibitem{FNP}
P. Facchi, H. Nakazato, and S. Pascazio,
Phys. Rev. Lett. {\bf 86}, 2699 (2001).

\bibitem{KS}
K. Koshino and A. Shimizu, 
Phys. Rev. A {\bf 67}, 042101 (2003).

\bibitem{Itano}
W. M. Itano, D. J. Heinzen, J. J. Bollinger, and D. J. Wineland,
Phys. Rev. A {\bf 41}, 2295 (1990).

\bibitem{Knight}
P. Knight, 
Nature {\bf 344}, 493 (1990). 

\bibitem{QZE_coll}
B. Nagels, L. J. F. Hermans, and P. L. Chapovsky, 
Phys. Rev. Lett. {\bf 79}, 3097 (1997).

\bibitem{laser_exp2}
Chr. Balzer, R. Huesmann, W. Neuhauser, and P. E. Toschek,
Opt. Commun. {\bf 180}, 115 (2000).

\bibitem{FGR}
M. C. Fischer, B. Gutierrez-Medina, and M. G. Raizen,
Phys. Rev. Lett. {\bf 87}, 040402 (2001).

\bibitem{NC}
M. A. Nielsen and I. L. Chuang,
{\it Quantum Computation and Quantum Information}
(Cambridge Univ. Press, 2000).

\bibitem{SF}
A. Shimizu and K. Fujita, {\it Proc. Quantum Control and Measurement '92}, 
(H. Ezawa and Y. Murayama, eds., Elsevier, 1993), p.191; quant-ph/9804026.

\bibitem{QNoise}
{\it Quantum Noise}, 
C. W. Gardiner, P. Zoller,
(Springer, Berlin, ed. 2, 2000).

\bibitem{Mandel}
{\it Optical Coherence and Quantum Optics},
L. Mandel and E. Wolf,
(Cambridge Univ. Press, Cambridge, 1995).

\bibitem{Gla}
R. J. Glauber, 
Phys. Rev. {\bf 130}, 2529 (1963). 

\bibitem{QND}
A. Shimizu, Phys. Rev. A {\bf 43}, 3819 (1991). 

\bibitem{finite-tauj}
Although we present the results only for $\tau_j \to 0$, 
a similar result is obtained for the case of 
finite $\tau_j$.

\bibitem{cm:finite-tau}
In fact, for unstable systems with finite $\tau_j$,
it has been confirmed in Ref.~\cite{KS} that
the present formalism completely
reproduces the projection-based theories in the limit of $\Delta\to\infty$.

\bibitem{CQED}
{\it Cavity Quantum Electrodynamics},
P. R. Berman, Ed.,
(Academic, San Diego, 1994).

\bibitem{sensitive}
In cavity QED systems,
the atomic decay is more sensitive to $\eta_{\veck}$ than the QZE systems
because of the large jump time $\tau_j$
due to the discreteness of the cavity mode.

\end{thebibliography}
\end{document}